\documentclass[preprint,superscriptaddress,showpacs]{revtex4}

\usepackage{graphics}
\usepackage{epsfig}

\begin{document}

\title{Hadron production in non linear relativistic mean field models}

\author{M. Chiapparini}\email{chiappa@uerj.br}
\affiliation{Instituto de F\'{\i}sica, Universidade do Estado do Rio
de Janeiro, \\ Rua S\~ao Francisco Xavier 524, Maracan\~a, CEP 20550-900
Rio de Janeiro, RJ, Brazil.}
\author{M. E. Bracco}
\affiliation{Faculdade de Tecnologia, Universidade do Estado do Rio 
de Janeiro, \\ Rodovia Presidente Dutra km 298, P\'olo Industrial, 
CEP 27537-000, Resende, RJ, Brazil.}
\author{A. Delfino}
\affiliation{Instituto de F\'{\i}sica, Universidade Federal Fluminense,\\
 Av. Gal. Milton Tavares de Souza s/n$^\circ$., Gragoat\'a, CEP 24210-346 
 Niter\'oi, RJ, Brazil.} 
\author{M. Malheiro}
\affiliation{Instituto Tecnol\'ogico da Aeron\'autica, 
Pra\c{c}a Marechal Eduardo Gomes 50, Vila das Ac\'acias, CEP 12228-900 
S\~ao Jos\'e dos Campos, SP, Brazil.}
\author{D. P. Menezes}
\affiliation{Depto de F\'{\i}sica, CFM, Universidade Federal de
Santa Catarina, \\ CP. 476, CEP 88040-900 Florian\'opolis, SC, Brazil.} 
\author{C. Provid\^encia}
\affiliation{Centro de F\'{\i}sica Computacional, Dep. de F\'{\i}sica, \\ 
Universidade de Coimbra, P-3004 - 516, Coimbra, Portugal.}

\begin{abstract}
By using a parametrization of the non-linear Walecka model which takes 
into account the binding energy of different hyperons, we present 
a study of particle production yields measured in central Au-Au collision 
at RHIC. Two sets of different hyperon-meson coupling constants are 
employed in obtaining the hadron production and chemical freeze-out parameters. 
These quantities show a weak dependence on the used hyperon-meson couplings. Results 
are in good overall accordance with experimental data. We have found  that the 
repulsion among the baryons is quite small and, through a preliminary 
analysis of the effective mesonic masses, we suggest a way to improve the fittings. 
\end{abstract}

\pacs{21.65.-f,24.10.Jv,25.75.-q}

\maketitle

\section{Introduction}
\label{intro}
Lattice simulations of QCD indicate that, at zero baryon number density and
temperatures of the order of 150-170 MeV, quarks and gluons become deconfined
\cite{fodor,karsch,petreczky}. This novel phase of nuclear matter is commonly
called quark-gluon plasma (QGP). It is believed that in nature temperatures 
of the order of
150 MeV existed only shortly after the Big Bang. For more than two decades,
attempts have been made to recreate similar conditions within the collision
of heavy nuclei at ultra-relativistic energies, looking for signatures of the
production of the QGP, which subsequently hadronizes. The hadronization stage
has been well discussed in the literature \cite{uheinz}. After this process
takes place, one assumes that a hadronic gas appears in a sizeable region
of space. Hydrodynamic models were used to describe this
hadronic evolution, based on the hypothesis that the fireball size is large
in comparison with the mean free path of the hadrons \cite{gbaym1,gbaym2}. One may ask
whether the hydrodynamic approach needs to consider viscosity and heat conduction 
effects, even before the freeze-out occurs. Here, one must say that the
employment of any kind of model has to be taken with caution.
In the gas regime, and for temperatures far from the system critical
temperature, the relaxation times which control the chemical equilibrium of
hadrons may become large compared with collision times, causing the gas to
reach the chemical equilibrium at an early stage of the expansion, favoring
an overpopulation of pions \cite{hleutwyler}. Temperature and chemical
potential are related to the chemical freeze-out regime. 
At low beam energies the chemical freeze-out of the resulting gas may be 
described 
by low temperature and high baryonic chemical potential. As the beam energy 
increases this situation reverts. The mesons become more numerous, the 
temperature 
increases and the chemical baryonic potential decreases.
For small baryonic chemical potential and high temperature the number of 
baryons becomes close
to the anti-baryons. Within this scenario, a QGP-hadronic matter phase 
transition is probably going to happen in heavy-ion collisions at 
very high energy, as the ones extected to take place soon 
at the Large Hadron Collider (LHC/ALICE).

Hadron multiplicities are observables which
can provide valuable information on the nature of the medium from which they
are produced. In this way, they are natural tools to look for QGP signatures.
Different models have been used to improve the understanding of the 
QGP-hadronic matter phase transition 
\cite{varios1,varios2,varios3,delfino1,delfino2}.
The use of thermal models to describe hadronic  
collision spectra has also been
considered \cite{brauna,braunb}. Curiously, despite the
fact that no microscopic theoretical basis was given for thermalization in 
such collisions, these models have been quite successful.

In thermal models, the statistical
distribution is controlled basically by the temperature and the chemical 
potential 
of a free hadron-gas. However, these ingredients alone does not suffice since
without a repulsion dynamics, the gas does not expand
properly. Therefore, for baryons and mesons, an excluded volume is needed
in the same spirit as in a Van der Waals gas. In particular, a detailed explanation 
of how incorporate a thermodynamically consistent excluded volume effect 
in a hadron gas was
presented in \cite{thermal1,thermal2}. The excluded volume parameters are
determined according to the fits of heavy-ion collision data.  
To fit Au-Au and Pb-Pb hadronic ratios data \cite{brauna,braunb,becatini}, a common
hard-core for baryons and mesons suffices.

In a different approach, that does not consider any correlation between baryons,
an interesting work \cite{cleymans} has shown that the energy per baryon
remains almost constant and equal to 1 GeV in a broad range of chemical
freeze-out temperatures.

More recently, studies including hadron-hadron dynamics in a sophisticated 
relativistic chiral SU(3) model, have shown that the hadron ratios obtained from
SIS, AGS, SPS to RHIC hadron energies are also well
described \cite{detlef}. This calculation, contrary to thermal models, takes into account
in medium hadronic mass effects. This new ingredient appears to modify
substantially the temperature and the baryonic chemical
potential to the freeze-out fittings. This happens even when the
SU(3) models are undergoing a first or a second order phase transition.
As a side information, let us remark that a first order phase
transition in a hadronic model induces a dramatic decrease in the
effective baryonic masses as well as a discontinuity in the entropy.
In the SU(3) model, both mesonic and baryonic masses are obtained with in medium
effects.

Between the simplicity of the thermal models \cite{brauna,braunb,becatini}
and the complexity of the relativistic chiral SU(3) model \cite{detlef},
one would ask whether well known relativistic
hadronic models \cite{walecka,serot} are able to successfully fit 
the mentioned chemical freeze-out hadronic ratios. In a previous study 
\cite{schaffner} the antibaryon $(\bar{p},\bar{\Lambda})$ production 
in relativistic nuclear collisions  using nonlinear models has been 
presented. The authors estimated the ratio of antibaryon and baryon 
densities in interacting and in  free systems. It was also pointed 
out that the annihilation of antibaryons in  surrounding matter at the final 
stage of the reaction may reduce their abundancy \cite{schaffner}. 
Relativistic hadronic models are based on field theories describing nuclear 
matter as a strongly
interacting system of baryons and mesons. This approach is 
assumed to be valid below the deconfinement phase
transition, and is based on the identification of 
the appropriate degrees of freedom at
this scale. The prototype of such theory is Quantum Hadrodynamics (QHD)
\cite{walecka,serot}, which has been shown to describe well many properties  
of nuclear matter, finite nuclei and neutron stars \cite{walecka,serot,horowitz,bogutaa,
glendenninga,gm,lalazissis}. This theory can be extended 
to include many-body correlations as density-dependent meson 
couplings \cite{tw,abcm}. 
The successful description of nuclear
properties indicates that the essential aspects of low energy strong
interactions are well described by QHD. At high energy, it can describe
dynamically the evolution from the hadronic fireball starting
scenario to the chemical freeze-out in which the baryons and the light 
mesons populate the gas. One of the advantages of using
relativistic quantum-hadronic models is that they can also describe the
already mentioned QGP-hadronic matter phase transition 
as the chemical potential increases \cite{delfino1,delfino2}.
The physics of the hadronic models at high temperature is very rich and 
its behavior has been well investigated for zero chemical potential \cite{theis}.  
Depending on their parametrization,
the models can present a first or second order phase transition in the same
way as the already mentioned SU(3) model for the hadronic masses. 

It is natural then to ask whether these routinely used
relativistic hadronic models can describe the hadron production
observed today in Au-Au collisions at RHIC \cite{braunb}. Recently, some
of these models, with and without density depend hadron-meson
coupling constants, have been used in order to answer that
question \cite{DCMMDM}. In that calculation not only the baryonic
octet was included but also the decuplet. A good overall description
of the particle yields data was obtained and in particular for
$K^{0*}/h^-$ and $\bar K^{0*}/h^-$ ratios, a good improvement was
achieved in comparison with the results obtained in the free gas
approximation \cite{braunb}. In \cite{DCMMDM} it was shown 
that the new dynamics achieved by the models with density depend parameters 
makes
the difference by having the best fits and that the mesons play an
important role. To go further with this investigation, we have chosen to 
consider in the present work only the baryonic octet because the inclusion of 
the decuplet produced only a small change in the overall results. Moreover,
we take into account the different binding energies of the hyperons 
\cite{bielich-gal,bielich,friedman-gal,dover-gal} in 
obtaining the chemical freeze-out parameters. We also
restrict ourselves to one parametrization of the non-linear Walecka
model \cite{walecka,serot,horowitz,bogutaa}.

In order to investigate the role played by the mesons, we have
introduced, in a very crude and naive way, an {\em ad-hoc} effective mesonic 
mass $m^*$ that increases in the medium and obeys a universal scaling with its 
bare value: $m^*/m=\alpha$, $\alpha>1$, for all mesons. Using a
best fitting analysis, we established a correlation between the
$\chi^{2}$ per degree of freedom ($\chi^2_{dof}$) and $\alpha$, understanding 
the result as an indication that the increase of 
the in-medium meson masses introduces additional repulsion among the
particles and mimics the repulsive effect obtained in other
approaches, where the hadron excluded volume is taken into account 
\cite{brauna,braunb}.

The paper is organized as follows. In Sec. II we present the models with a 
detailed discussion on chemical, strangeness and charge equilibrium. In Sec.
III, we present the parametrizations used for the couplings between hyperons 
and mesons. In Sec. IV, we present our results and summarize our main 
conclusions.

\section{The model}
\label{models}
We assume that the chemical freeze-out can be described as a mixture of
the lightest baryons and mesons. We use the following Lagrangian
\begin{equation}
{\cal L}={\cal L}_{QHD}+{\cal L}_{\pi K\rho},\label{lag1}
\end{equation}
where
\begin{equation}
{\cal L}_{QHD}={\cal L}_b+{\cal L}_m,\label{lag2}
\end{equation}
with
\begin{eqnarray}
{\cal L}_b&=&
 \sum_{j} \bar{\psi}_j \left\{ \gamma_\mu \left[i\partial^\mu
  -g_{\omega j}\omega^\mu - g_{\rho j} I_{3j} \rho^{0\mu}\right]
  -\left(m_j-g_{\sigma j}\sigma\right)\right\}\psi_j ,  \label{lag21}\\
{\cal L}_m & = & \frac{1}{2}\left(\partial_\mu\sigma\partial^\mu\sigma -
m^2_\sigma\sigma^2\right)
-\frac{b}{3}\left(g_{\sigma N}\sigma\right)^3
-\frac{c}{4}\left(g_{\sigma N}\sigma\right)^4 \nonumber
-\frac{1}{4}\left(\partial_\mu\omega_\nu-\partial_\nu\omega_\mu \right)^2 \label{lag22}\\
&&+\frac{1}{2}m^2_\omega\omega_\mu\omega^\mu \nonumber
-\frac{1}{4}\left(\partial_\mu\rho^0_\nu-
\partial_\nu\rho^0_\mu\right)^2+
\frac{1}{2}m^2_\rho \rho^0_\mu \rho^{0\mu}.
\end{eqnarray}
Mesons $\{\pi^\pm,K^0,\bar{K^0},K^\pm,\rho^\pm,{K^*}^0,\bar{{K^*}^0},{K^*}^\pm\}$
enter Lagrangian ${\cal L}_{\pi K\rho}$ in (\ref{lag1}) as a free gas of bosons.
 Lagrangian
(\ref{lag2}) has a general form commonly used in relativistic mean field models. The sum over
$j$ in (\ref{lag21}) extends over the octet of lightest baryons
$\{n,p,\Lambda,\Sigma^-,\Sigma^0,\Sigma^+,\Xi^-,\Xi^0\}$, and their
antiparticles. $I_{3j}$ is the corresponding isospin quantum number. The 
baryon and meson masses are listed in Table \ref{masses}. The values for the couplings 
are taken from \cite{gm} and shown in Table \ref{param}.
\begin{table}[pt]
\begin{center}
\begin{tabular}{ccccccccc} \\
\hline
           & $N$ & $\Lambda$ & $\Sigma$ & $\Xi$ & $\pi$ & $K$ & $\rho$ &$K^*$\\ \hline
$m$ (MeV)  & 939 &  1116     &  1193    &  1318 &  138  & 495 & 776    &893  \\
\hline
\end{tabular}
\caption{Masses used in the calculation.}
\label{masses}
\end{center}
\end{table}
\begin{table}[pt]
\begin{center}
\begin{tabular}{ccccccc} \\
\hline
$K$ (MeV)& $m^*/m$ & $l_\sigma$ (fm$^{-2}$)&$l_\omega$ (fm$^{-2}$)&$l_\rho$ (fm$^{-2}$)&$b$ (fm$^{-1}$)&$c$\\ \hline
300      & 0.70    &   11.79               & 7.149                & 4.411              & 0.01402       &-0.001070  \\
\hline
\end{tabular}
\caption{Parameters used in the calculation ($l_i=g_{iN}^2/m_i^2$), taken from Ref. \protect\cite{gm} ($b$ here includes the product with $m_N$ in Eq. (1) of this reference).}
\label{param}
\end{center}
\end{table}
The Euler-Lagrange equations of motion for the mediating mesons in a system 
with translational and  rotational invariance, within
the mean field approximation, are
\begin{eqnarray}
V_\sigma&=& l_\sigma\left(\sum_{j}x_{\sigma j}n_{sj}-bV_\sigma^2-{\rm c}V_\sigma^3\right), \label{vsigma} \\
V_\omega&=& l_\omega\sum_{j}x_{\omega j}n_{j}, \label{vomega}\\
V_\rho&=& l_\rho\sum_{j}x_{\rho j}n_{3j}, \label{vrho}
\end{eqnarray}
with $V_\sigma=g_{\sigma N} \sigma$, $V_\omega=g_{\omega N} \omega_0$, $V_\rho=g_{\rho N} \rho^0_0$, $x_{ij}=g_{ij}/g_{iN}$ and $l_i=g_{iN}^2/m_i^2,\;\;i=\sigma,\omega,\rho$. The densities entering the above equations are
\begin{eqnarray}
n_{sj}&=& \frac{1}{\pi^2}\int  \frac{m_j^*}{\epsilon_j(p)}(f_{j+}+f_{j-})p^2dp,\\
n_{j}&=& \frac{1}{\pi^2}\int (f_{j+}-f_{j-})p^2dp,\\
n_{3j}&=&I_{3j}n_j,
\end{eqnarray}
where $m_j^*=m_j-g_{\sigma j}\sigma$ is the effective mass of the baryon $j$, and $\epsilon_j(p)=\sqrt{p^2+{m_j^*}^2}$. The baryonic distribution functions $f_{j\pm}$ are the Fermi-Dirac ones
\begin{equation}
f_{j\pm}=\frac{1}{\exp{\left[(\epsilon_j\mp\nu_j)/T\right]}+1},
\end{equation}
where the sign minus (plus) accounts for particles (antiparticles). The baryonic effective chemical potential $\nu_j$ is defined as
\begin{equation}
\nu_j=\mu_j-g_{\omega j}\omega_0-g_{\rho j}I_{3j}\rho_0^0,
\end{equation}
being $\mu_j$ the thermodynamical chemical potential of baryon $j$. The particle density of the free mesons $\pi$, $K$, $\rho$ and $K^*$ are calculated using the Bose-Einstein distribution function in the form
\begin{equation}
n_l=\frac{\gamma_l}{2\pi^2}\int\frac{1}{\exp{\left[(\epsilon_l-\mu_l)/T\right]}-1}p^2dp,\;\;\;l=\pi,K,\rho,K^*
\end{equation}
where $\epsilon_l=\sqrt{p^2+m_l^2}\,$, $\mu_l$ is the chemical potential, and $\gamma_l$ is the corresponding spin degeneracy.

The problem is solved in the grand canonical ensemble, imposing the constraints of baryon number, strangeness and electric charge conservation, given respectively by \cite{brauna}:
\begin{eqnarray}
V\sum_in_iB_i&=&Q_B=Z+N, \label{cob}\\
V\sum_in_iS_i&=&Q_S=0, \label{cos}\\
V\sum_in_iI_{3i}&=&Q_{I_3}=(Z-N)/2, \label{coi3}
\end{eqnarray}
where $i$ runs over all particles in the system, and $\{B_i,S_i,I_{3i}\}$ are the baryonic, strangeness and isospin quantum numbers of particle $i$. 
Each conserved charge has a conjugated  
chemical potential, namely $\{Q_B,Q_S,Q_{I_3}\}\leftrightarrow\{\mu_B,\mu_S,\mu_{I_3}\}$. The temperature $T$ and the baryochemical potential $\mu_B$ are the two independent parameters of the model, while the 
volume of the fireball $V$, the strangeness chemical potential $\mu_S$, and the isospin chemical potential $\mu_{I_3}$ are fixed by the three constraints (\ref{cob})-(\ref{coi3}). The chemical potential of each 
particle in the system can be written as a linear combination of these three conjugate chemical potentials in the form
\begin{equation}
\mu_i=B_i\mu_B+I_{3i}\mu_{I_3}+S_i\mu_S. \label{mug}
\end{equation}

In the baryonic sector, a zero chemical potential implies a zero baryonic
density since the number of
baryons is equal to the anti-baryons. For the $\pi^0$, for instance, zero
chemical potential is obtained directly from Eq. (\ref{mug}), using 
$B_{\pi^0}=I_{3\pi^0}=S_{\pi^0}=0$. Despite the fact that $\mu_{\pi^0}=0$,
the density for $\pi^0$ is different from zero. This remark  
applies only to $\pi^0$ and $\rho^0$. In particular, $\pi^0$ 
results completely decoupled from the other particles. Its population is that 
of a boson gas at temperature $T$ and zero chemical potential. This is
the reason why it was not considered in Lagrangian ${\cal L}_{\pi K\rho}$ in 
(\ref{lag1}). Consistently, particle ratios do not 
involve the $\pi^0$ and $\rho^0$ mesons \cite{braunb}. Hence, 
these particles do not need to be considered in the equations of the chemical potentials. 
Note that the $\rho^0$ meson appears in Lagrangian ${\cal L}_m$, 
Eq. (\ref{lag21}), only as a mediating field in the mean field approximation. 

In the bosonic sector, the chemical potentials have to obey the physical
inequality,
\begin{equation}
\mu_l \leq m_l, \label{mum}
\end{equation}
where $m_l$ is the bosonic mass.
This constraint is imposed naturally by statistical mechanics in order to 
keep the bosonic partition function finite.

The constraints contained in Eqs.(\ref{mug},\ref{mum}) allow us to restrict 
the possible values of the generalized chemical potentials by
 the following inequalities:
\begin{equation}
0\leq |\mu_{I_3}| \leq m_{\pi}, \label{ineq1}
\end{equation}
\begin{equation}
0\leq |\mu_{S}| \leq -\frac{1}{2}|\mu_{I_3}|+m_K .\label{ineq2}
\end{equation}

This set of inequalities was implemented numerically in order to solve the 
large self-consistent set of equations of the problem.

\section{The hyperonic couplings}
\label{hcouplings}

The coupling constants of the hyperons to the $\omega$ and $\rho$ mesons 
are fixed using SU(6) symmetry \cite{dover-gal}, which means that the vector 
coupling constants scale with the number of light quarks in the baryon in the form $g_{\omega N}:g_{\omega\Lambda}:g_{\omega\Sigma}:g_{\omega\Xi}=3:2:2:1$. The ratios of the vector couplings are then fixed to
\begin{equation}
x_{\omega\Lambda}=x_{\omega\Sigma}=\frac{2}{3}, \;\;\;\;\;\; 
x_{\omega\Xi}=\frac{1}{3},\label{gvh}
\end{equation}
meaning that the vector coupling constants of the hyperonic sector are fixed once the $g_{\omega N}$ coupling constant is know. 

The coupling strength of the $\rho$ meson, however, is given by the isospin of the baryon in the form $g_{\rho N}:g_{\rho\Lambda}:g_{\rho\Sigma}:g_{\rho\Xi}=1/2:0:1:1/2$. Again, the isovector coupling constants of the hyperonic sector are fixed once $g_{\rho N}$ is know. This is taken into account automatically with the specific form of the isovector couplings in Lagrangian (\ref{lag21}) choosing
\begin{equation}
x_{\rho i}=1, \;\;\;\;\;\;\;\;i=\Lambda,\Sigma,\Xi. \label{grh}
\end{equation} 
The coupling constants $\{x_{\sigma i}\}_{i=\Lambda,\Sigma,\Xi}$ of the hyperons 
with the scalar meson $\sigma$ 
are constrained by the hypernuclear potentials in nuclear matter to be 
consistent with hypernuclear data \cite{bielich-gal,bielich,friedman-gal}. 
The hypernuclear potentials were constructed as
\begin{equation}
V_i=x_{\omega i}V_{\omega N} - x_{\sigma i}V_{\sigma N}, \;\;\;\;\;\;\;\;i=\Lambda,\Sigma,\Xi, \label{vh1}
\end{equation}
where $V_{\omega N}=215.83$ MeV and $V_{\sigma N}=281.47$ MeV are the nuclear potentials for symmetric nuclear matter at saturation with the parameters of Table \ref{param}. Following Ref. \cite{bielich} we use 
\begin{equation}
V_\Lambda=-28\;\mbox{MeV},\;\;\;\;\;\; 
V_\Sigma = 30\;\mbox{MeV},\;\;\;\;\;\;
V_\Xi= -18\;\mbox{MeV}. \label{vh2}
\end{equation}
All hyperon coupling ratios $\{x_{\sigma i},x_{\omega i},x_{\rho i}\}_{i=\Lambda,\Sigma,\Xi}$ are now known once the coupling constants $\{g_{\sigma N},g_{\omega N},g_{\rho N}\}$ of the nucleon sector are given.

We have to guarantee also that the hyperon coupling ratios are constrained 
to the known neutron star maximum masses, as in Ref. \cite{gm}. In this reference windows are established for the values of $x_{\sigma}$ and 
$x_{\omega}$: $x_{\sigma}$ and $x_{\omega}$ are the couplings between 
hyperons and $\sigma$ and $\omega$ meson respectively and they have
to be compatible with the allowed maximum neutron star masses and with the $\Lambda$ binding 
energy in nuclear matter. Couplings $x_\sigma$ and $x_\omega$
are the same for all hyperons and the $x_\rho$ couplings are fixed arbitrarily as $x_\rho=x_\sigma$. The SU(6) value $x_\omega=2/3$ of Eq.(\ref{gvh}) together with $x_\sigma=0.6104$, is one of the allowed couple of values (they were interpolated from Table I of Ref. \cite{gm}). This choice corresponds to  Set1 row in Table \ref{xs}. The other couplings used in Ref. \cite{gm} are those shown in Table \ref{param}. Through this work we refer to this relativistic model as GM1.
\begin{table}[pt]
\begin{center}
\begin{tabular}{cccccccc} \\
\hline
     &$x_{\omega\Lambda}$&$x_{\omega\Sigma}$&$x_{\omega\Xi}$&$x_{\sigma\Lambda}$&$x_{\sigma\Sigma}$&$x_{\sigma\Xi}$&$x_{\rho i}$  \\ \hline
Set1&     0.6666        &       0.6666     &      0.6666    &       0.6104      &       0.6104     &      0.6104  & 0.6104\\
Set2 &     0.6666        &       0.6666     &      0.3333    &       0.6106      &       0.4046     &      0.3195  &  1.000\\
\hline
\end{tabular}
\caption{Parametrizations used for the hyperon coupling constants.
Set1 is taken from Ref. \protect\cite{gm}. Set2  is obtained from Set1 
imposing SU(6) symmetry and different binding energies for hyperons in nuclear matter 
(see text for details). The last column, $x_{\rho i}$, refers to the coupling of the $\rho$ meson with all hyperons, $i=\Lambda,\Sigma,\Xi$.}
\label{xs}
\end{center}
\end{table}
Now we modify Set1 to take into account constraints (\ref{gvh})-(\ref{vh2}). 
The value $x_{\omega\Lambda}=2/3$ is kept constant and 
the others are modified accordingly with the constraints. The resulting set of hyperonic couplings is the Set2 row in Table \ref{xs}. 

\begin{figure}
\begin{center}
\includegraphics[width=8cm]{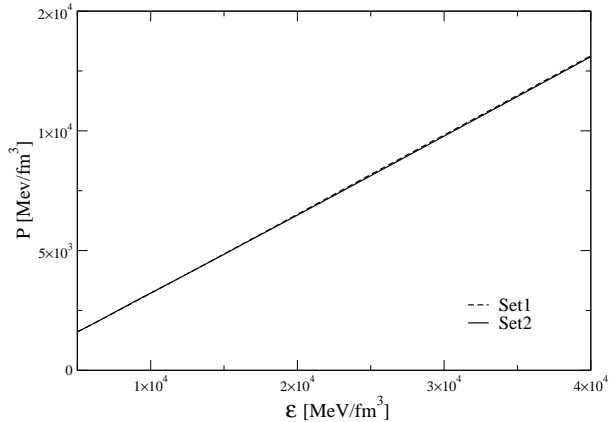}
\end{center}
\caption{Equations of state for the relativistic models of 
Ref. \protect\cite{gm} (dashed line, Set1 in Table \protect\ref{xs}) and the one of this work (full line, Set2 in Table \protect\ref{xs}). 
Both equations of state are almost indistinguishable, ensuring the same neutron-star maximun masses.}
\label{ede}
\end{figure}

To see the impact of this redefinition of hyperonic couplings on 
neutron star properties, we show in Fig. \ref{ede} the equations of state calculated using the model of Ref. \cite{gm} with Set1 (dashed line) and 
with Set2  (full line) of Table \ref{xs} for the hyperon coupling constants.  
As we can see, both curves are almost indistinguishable, ensuring the same neutron star properties. This validates Set2  as a set of 
hyperonic couplings that preserves maximum neutron star masses as well as experimental results for the binding energy of hyperons in symmetric nuclear matter.
In the following, we use Set1 and Set2  for the hyperonic couplings in 
the Lagrangian (\ref{lag1}) together with parameters of Table \ref{param} 
to investigate particle production in relativistic heavy ion collisions. 

\section{Results and conclusions}
\label{results}

We apply the above formalism to the description of the experimental
data for hadron production yield in Au-Au collisions at RHIC from
STAR, PHENIX, PHOBOS and BRAHMS collaborations, following
\cite{braunb} and references therein. We adjust the free parameters
$T$ and $\mu_B$ to get the best description of the data, based on a
$\chi^2$ analysis of the form \cite{brauna} 
\begin{equation}
\chi^2=\sum_i\left(\frac{R^{exp.}_i-R^{mod.}_i}{\sigma_i}\right)^2,
\end{equation}
where $R^{mod.}_i$ is the ith particle ratio obtained from the models used 
here and $\sigma_i$ is the experimental error. When more than one 
experimental data is available, we calculated a weighted average. This 
leaves us with seven effective degrees of freedom for the fit.

Our results are summarized in Table \ref{ratios} for the hyperon coupling
constants given in Table \ref{xs}. 
\begin{table}[pt]
\begin{center}
\begin{tabular}{cccccccc} \\
\hline
                       &\multicolumn{2}{c}{GM1}& Free  & Free$^*$ &Th    &Exp. Data&Exp. \\
Ratio                  &Set1         &Set2       &     &          &               &     \\ \hline

$\bar{p}/p$            &0.673        &0.676      &0.674 & 0.638    &0.629 &0.65$\pm$0.07 & STAR  \\
                       &             &           &      &          &      &0.64$\pm$0.07 & PHENIX\\
                       &             &           &      &          &      &0.60$\pm$0.07 & PHOBOS\\
                       &             &           &      &          &      &0.64$\pm$0.07 & BRAHMS\\
$\bar{p}/\pi^-$        &0.048        &0.046      &0.038 &0.080     &0.078 &0.08$\pm$0.01 & STAR  \\
$\pi^-/\pi^+$          &1.005        &1.005      &1.004 &1.011     &1.007 &1.00$\pm$0.02 & PHOBOS\\
                       &             &           &      &          &      &0.95$\pm$0.06 & BRAHMS\\
$K^-/K^+$              &0.957        &0.959      &0.964 &0.888     &0.894 &0.88$\pm$0.05 & STAR  \\
                       &             &           &      &          &      &0.78$\pm$0.13 & PHENIX\\
                       &             &           &      &          &      &0.91$\pm$0.09 & PHOBOS\\
                       &             &           &      &          &      &0.89$\pm$0.07 & BRAHMS\\
$K^-/\pi^-$            &0.239        &0.237      &0.231 &0.175     &0.145 &0.149$\pm$0.02& STAR  \\
${K^*}^0/h^-$          &0.063        &0.062      &0.059 &0.038     &0.037 &0.06$\pm$0.017& STAR  \\
$\bar{{K^*}^0}/h^-$    &0.060        &0.059      &0.057 &0.033     &0.032 &0.058$\pm$0.017& STAR \\
$\bar{\Lambda}/\Lambda$&0.692        &0.694      &0.699 &0.717     &0.753 &0.77$\pm$0.07 & STAR  \\
$\Xi^+/\Xi^-$          &0.723        &0.713      &0.725 &0.806     &0.894 &0.82$\pm$0.08 & STAR  \\
\hline
$T$ (MeV)              &148.6        &148.0      &145.7 &169.4     &174  &              &       \\
$\mu_B$ (MeV)          &33.27        &32.51      &28.88 &38.52     &46   &              &       \\
$\chi^2_{dof}$         &5.65         &5.77       &6.24  &1.08      &0.81 &              &       \\
\hline
\end{tabular}
\caption{Comparisons of experimental particle ratios with the ones obtained
from the relativistic mean-field models used in this work for the hyperon 
couplings belonging to Set1 and Set2 , together with the chemical
freeze-out temperature $T$, baryonic potential $\mu_B$ and $\chi^2_{dof}$. Also shown are
the results from the free gas (Free), from the free gas with {\em ad-hoc} effective meson masses (Free$^*$, $m^*/m=1.3$), and from the thermal model of 
\cite{braunb} (Th). Experimental values of Au-Au 
collisions at $\sqrt{s}=130$ MeV were taken from that reference and references therein.}
\label{ratios}
\end{center}
\end{table}
We also show, in the fourth column, the results from a non-interacting gas 
of baryons and mesons (Free), in the fifth column the results from a non-interacting 
gas of baryons and mesons with effective meson masses as discussed below (Free$^*$), 
together with the results from the thermal model of Ref. \cite{braunb} in column six (Th). 

We can see that both set of hyperonic couplings give almost the same 
results for the particle ratios. They are not very different from the model with zero coupling constants designated by Free, although in this case a smaller temperature and chemical potential were obtained. As a whole the fractions are well described, but we observe that there is a systematic deviation for the fractions involving mesons except for $\pi^-/\pi^+$. This reflects the naive way the mesons were included. However, we point out that the ${K^*}^0/h^-$ and $\bar{{K^*}^0}/h^-$ ratios have improved with respect to the thermal model results in accordance with our previous work \cite{DCMMDM} and also with the chiral SU(3) calculations \cite{detlef}. For $h^-$ we designated all negatively charged hadrons produced, including mesons and baryons.

Regarding the quality of the fit, the resulting $\chi^{2}_{dof}$ for both 
sets are around 5.70, showing a slight improvement of Set1 over Set2 . When  
these values are multiplied  by the seven degrees of freedom used in 
our fitting, $\chi^{2}$ reaches 40, which is a high value.  
This does not happen with the fits
obtained by thermal models \cite{brauna,braunb,becatini} and with
the chiral SU(3) model \cite{detlef}, for which $\chi^{2}$ is around 6. 

Looking now at the freeze-out temperatures, we see that they are 
almost the same for the two sets of hyperonic couplings considered  
($T = 148.6-148.0$ MeV), showing a small decrease when going from 
Set1 to Set2 .  This situation repeats for the chemical potential fitting 
($\mu = 33.27-32.51$ MeV). It is important to mention at this point 
that in these models the hadronic phase transition regions occur at 
temperatures $T>180$ MeV \cite{delfino2}, higher than the freeze-out 
temperature we found here.

In order to better understand why we obtain a high
$\chi^{2}_{dof}$, we argue that the repulsion among the
baryons is quite small and cannot describe so well the chemical
freeze-out environment. This happens because the chemical potential is too
low in comparison with the temperature. The other cited approaches certainly have
sufficient built in repulsion to avoid an overpopulation of
particles. In the thermal model, it is done through excluded
volumes introduced in a thermodynamically consistent way. On the
other hand, the chiral SU(3) model  includes meson-meson and
meson-baryon interactions, providing effective masses for all
hadrons, giving a higher kinetic energy which mimics an additional
repulsion among the particles.

We think that the relativistic models themselves suffer from a
lack of repulsive correlations at low densities, where the sources for 
the mediating meson fields are very weak. For example, in the GM1 model 
with Set2 , studied in this work, we have $V_{\sigma N}=21.44$ MeV, $V_{\omega N}=3.359$ MeV, 
and $V_{\rho N}=-0.8919\times 10^{-2}$ MeV.
This fact results in a system almost identical to the one described by
a free gas of baryons and mesons, as we can see comparing columns 
2, 3 and 4 in Table \ref{ratios}. Therefore, the hadron production 
is not very sensitive to the hadron-meson
interactions, manifested in the almost model independence of the 
$\chi^{2}_{dof}$ obtained and also in the small dependence of the
freeze-out parameters. In spite of the reduced number of models
studied here, the results obtained in \cite{DCMMDM}, where
more models were considered, also reinforce this point. This conjecture 
may be extended to all relativistic hadronic models commonly used 
in the literature.

Understanding the reason underlying the lack of repulsion in the 
description of the chemical freeze-out regime is a relevant question.
 In the following, we present a crude argument for that.
 Let us suppose, for simplicity,  that baryon-baryon interaction proceeds 
 through an attractive scalar ($S$) and a repulsive vector ($V$) meson-exchange
potentials. In the mean field approach, at zero
temperature and at normal saturated nuclear matter, the
depth of $\Sigma=S+V$ is around $60$ MeV. The temperature increase
in these models favors the appearance of anti-baryons. Table \ref{ratios} 
 shows that the $\bar p  /p $ ratio is around 0.67.
 However, from G-parity symmetry, the anti-baryons
themselves revert the sign of the vector potential in such a way
that $\Sigma=S-V$ \cite{mishustin,greiner,mishustin2}. Thus, anti-baryon-baryon
interactions themselves do not carry repulsion. Therefore, the
baryon-baryon repulsion alone has to compensate for the large attraction
among the anti-baryons, explaining the lack of repulsion in the
hadronic models.

\begin{figure}
\begin{center}
\includegraphics[width=8cm]{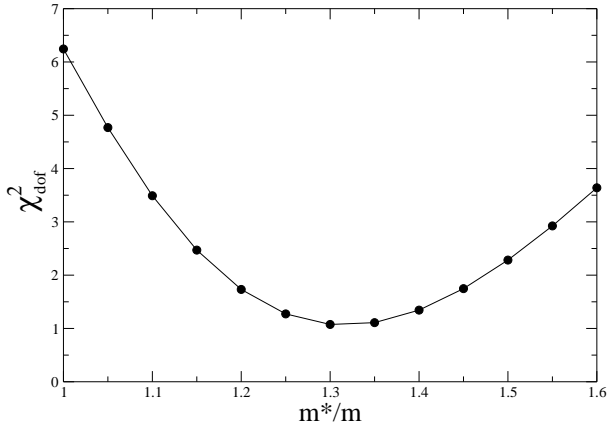}
\end{center}
\caption{$\chi^2_{dof}$ as a function of $m^*/m$ (the same for all mesons),
where $m$ is the meson bare mass, for a free relativistic gas of baryons and mesons (baryon masses are mantained fixed to the experimental values of Table \ref{masses}).}
\label{m-chi2}
\end{figure}

One way of implementing a repulsive content in this class of hadronic
models would be by introducing an excluded volume for the particles
in the system \cite{thermal1,thermal2,ev}. Of course, this approach requires the recalculation
of the coupling constants in order to maintain the correct normal
nuclear matter properties of the original models. By doing
so, however, the hadronic models would loose its strong appealing
due to their theoretical basis together with the success in
reproducing so many observables in several different regimes of nuclear
matter.  Another possibility, inspired by the SU(3) hadronic model
\cite{detlef}, is to include meson-meson self-coupling and meson-baryon
couplings for the heavy mesonic sector, as $K$ and $\rho$ for example. 
We have studied this effect in an {\em ad-hoc} and crude way. In the
statistical thermal distributions of the free gas, we have artificially 
assumed in-medium effective mesonic masses $m^*$, instead of the bare ones in vacuum. In this preliminary study, we have used the same value of
$m^*/m$ for all mesons. In Figure \ref{m-chi2} we see how $\chi^{2}_{dof}$ becomes smaller as $m^*/m$ increases from 1, reaches a minimum around $m^*/m\approx 1.3$, and starts to grow again. 
This shows that an increase in the
effective in-medium mesonic mass goes in the right
direction towards obtaining a better fit of the data. This is confirmed 
looking at column 5 in Table \ref{ratios}, where the particle ratios for 
$m^*/m=1.3$, close to the minimum of Fig. \ref{m-chi2}, are shown. The increase 
of the in-medium meson masses would be compatible with the notion of a 
thermal mass which increases linearly with temperature. This kind of 
behaviour can be seen in chiral effective models \cite{detlef} and also in the Walecka model \cite{mcj93}. Including the effective meson masses in this {\em ad-hoc} way, we automatically improve the ratios involving mesons, and reduce the fractions involving $h^-$. We point out, however, that in fact the uncertainty on the measured values of ${K^*}^0/h^-$ and ${\bar{K^*}^0}/h^-$ are quite large. Also, the equilibrium temperatures increases more than 15\% and the chemical potential 33\%, coming closer to the results of the thermal model.

This finding strongly suggests that introducing an appropriate dynamic 
treatment of the mesons,
 allowing for the modification of their masses in medium, would favor
 the correct data ratios without the need of including excluded volumes.
However, the model dependence addressed in Ref. \cite{detlef} remains to
be further investigated. 

Another remark regards the particle ratios in the medium as we have 
calculated and the ones measured. The later comes from asymptotic particle 
states with vaccuum masses, i.e., on the mass shell $(m^*=m)$. In our 
model calculation the particles are off-shell and any mechanism to bring 
them on-shell would modify the ratios. However, we did not include annihilation 
and nonequilibrium processes which are claimed to reduce the calculated 
yields \cite{schaffner}.

Summarizing, in this work we have investigated the
effects of different hyperon-meson coupling constants in hadron production and 
chemical freeze-out parameters. We have observed that the freeze-out 
temperatures and chemical potentials show a small dependence on the different parametrizations. Our results indicate that the repulsion among 
the baryons is quite small and the hadron production is not very sensitive to 
the hadron-meson interactions. A simple {\em ad hoc} in-medium mesonic mass 
study shows the importance of having meson-meson interaction in the hadronic models.

\section*{Acknowledgments}

This work was partially supported by CNPq (Brazil) and FEDER/FCT (Portugal)
under the project PTDC/FIS/64707/2006. Two of the authors 
(M.C. and M.E.B.) would like to thank Prof. Marcelo Munhoz for valuable 
discussions.

\end{document}